\documentclass[10pt,
aps,
prl,
twocolumn,
amsmath,
amssymb,
footinbib,
showpacs,
longbibliography,
superscriptaddress,
floatfix,
]{revtex4-2}

\usepackage[utf8]{inputenc}
\usepackage[T1]{fontenc}

\usepackage{newtxtext,newtxmath}

\usepackage[normalem]{ulem}
\usepackage[english]{babel}
\usepackage{latexsym}
\usepackage{subfigure}
\usepackage{epsfig}
\usepackage{color}
\usepackage{hyperref}
\usepackage[capitalize]{cleveref}
\usepackage{amstext}\usepackage{amssymb}\usepackage{amsmath}\usepackage{mathtools}
\usepackage{graphicx}
\usepackage{tikz}
\usepackage{esint}
\usepackage{physics}
\usepackage{soul}
\hypersetup{
colorlinks=true,
citecolor=blue,
linkcolor=red,
urlcolor=black
}
\AtBeginDocument{%
    \newwrite\bibnotes
    \def\bibnotesext{Notes2.bib}
    \immediate\openout\bibnotes=\jobname\bibnotesext
    \immediate\write\bibnotes{@CONTROL{REVTEX41Control}}
    \immediate\write\bibnotes{@CONTROL{%
    apsrev41Control,author="08",editor="1",pages="1",title="0",year="1"}}
     \if@filesw
     \immediate\write\@auxout{\string\citation{apsrev41Control}}%
    \fi
}%

\definecolor{darkblue}{rgb}{0,0,0.7}
\definecolor{darkred}{rgb}{0.7,0,0}
\definecolor{darkgreen}{rgb}{0,0.4,0}
\definecolor{orange}{rgb}{0.8,0.4,0}

\newcommand{\ie}{\textit{i.e.\@}}

\newcommand{\betaCE}{\beta_{\textrm{\tiny CE}}}
\newcommand{\betaSWT}{\beta_{\textrm{\tiny LSWT}}}

\newcommand{\betam}{\beta_{\textrm{\tiny m}}}

\newcommand{\betaEE}[1]{\beta_{\textrm{\tiny EE}}^{#1}}

\newcommand{\betaSE}{\beta_{\textrm{\tiny SE}}}

\newcommand{\Vg}{V_{\textrm{g}}}
\newcommand{\Vphi}{V_{\varphi}}

\newcommand{\VCE}{V_{\textrm{\tiny CE}}}
\newcommand{\VSE}{V_{\textrm{\tiny SE}}}

\newcommand{\Vm}{V_{\textrm{m}}}

\newcommand{\lettersection}[1]{\paragraph*{#1.---}}
\renewcommand{\acknowledgments}{\bigskip\noindent}


\begin{document}

\title{Spreading of Correlations and Entanglement in the Long-Range Transverse Ising Chain}

\author{J.~T.~Schneider}
\affiliation{CPHT, CNRS, Ecole Polytechnique, IP Paris, F-91128 Palaiseau, France}

\author{J.~Despres}
\affiliation{CPHT, CNRS, Ecole Polytechnique, IP Paris, F-91128 Palaiseau, France}

\author{S.~J.~Thomson}
\affiliation{CPHT, CNRS, Ecole Polytechnique, IP Paris, F-91128 Palaiseau, France}

\author{L.~Tagliacozzo}
\affiliation{Department of Physics and SUPA, University of Strathclyde, Glasgow G4 0NG, United Kingdom}
\affiliation{Departament de Física Quàntica i Astrofísica  and Institut de Ciències del Cosmos (ICCUB), Universitat de Barcelona, Martí i Franquès 1, 08028 Barcelona, Catalonia, Spain}

\author{L.~Sanchez-Palencia}
\affiliation{CPHT, CNRS, Ecole Polytechnique, IP Paris, F-91128 Palaiseau, France}

\date{5 March 2021}

\begin{abstract}
  Whether long-range interactions allow for a form of causality in non-relativistic quantum models remains an open question with far-reaching implications for the propagation of information and thermalization processes.
  Here, we study the out-of-equilibrium dynamics of the one-dimensional transverse Ising model with algebraic long-range exchange coupling. Using a state of the art tensor-network approach, complemented by analytic calculations and considering various observables, we show that a weak form of causality emerges, characterized by non-universal dynamical exponents.
  While the local spin and spin correlation causal edges are sub-ballistic, the causal region has a rich internal structure, which, depending on the observable, displays ballistic or super-ballistic features.
  In contrast, the causal region of entanglement entropy is featureless and its edge is always ballistic, irrespective of the interaction range.
  Our results shed light on the propagation of information in long-range interacting lattice models and pave the way to future experiments, which are discussed.
\end{abstract}

\maketitle

Long-range interactions may dramatically impact the dynamics of correlated systems~\cite{dauxois2002}.
In the quantum regime, a number of basic concepts, such as the equivalence of the thermodynamic ensembles~\cite{kastner2010a,*kastner2010b}, the Mermin-Wagner-Hohenberg theorem~\cite{mermin1966,hohenberg1967,peter2012} or the area law for entanglement entropy~\cite{Eisert+08,koffel2012,cadarso2013,schachenmayer2013}, break down.
Long-range interactions may also be responsible for negative heat capacities and anomalous response functions~\cite{thirring1970}.
The paradigmatic model considers interactions that fall off algebraically with the distance \(R\), as \(1/R^\alpha\).
Such long-range interactions can now be emulated in quantum simulators~\cite{buluta2009,georgescu2014,NaturePhysicsInsight2012cirac, *NaturePhysicsInsight2012bloch,*NaturePhysicsInsight2012blatt,*NaturePhysicsInsight2012aspuru-guzik,*NaturePhysicsInsight2012houck,gross2017,lsp2018,*tarruell2018,*aidelsburger2018,*lebreuilly2018,*LeHur2018,*bell2018,*alet2018} with
artificial ion crystals~\cite{deng2005,islam2011,lanyon2011,schneider2012b,NaturePhysicsInsight2012blatt,bermudez2017},
Rydberg gases~\cite{bendkowsky2009,weimer2010,schausz2012,browaeys2016},
magnetic atoms ~\cite{griesmaier2005,beaufils2008,lu2011,baier2016,lahaye2009},
polar molecules~\cite{micheli2006,yan2013,moses2017},
nonlinear optical media~\cite{firstenberg2013},
and solid-state defects~\cite{childress2006,balasubramanian2009,dolde2013}.
A major asset of these systems is that they are free of screening effects and the interaction is truly long range.
Moreover, the exponent \(\alpha\) can be controlled~\cite{deng2005,browaeys2016}.

Another dramatic effect of long-range interactions is the breakdown of the notion of causality in non-relativistic quantum models.
For a wide class of lattice models with short-range interactions, the Lieb-Robinson bound implies that correlations decay exponentially beyond a limit set by the system's maximum group velocity~\cite{lieb1972,hastings2006,calabrese2006}.
This effective light-cone effect has been demonstrated for various models in both experiments~\cite{cheneau2012,fukuhara2013,geiger2014} and numerics~\cite{lauchli2008,manmana2009,barmettler2012,carleo2014,despres2019,Villa+19,Villa+20}.
In contrast, for sufficiently long-range interactions, causality breaks down and information can propagate arbitrarily fast~\cite{hauke2013,eisert2013}. This is consistent with the absence of known Lieb-Robinson bounds~\cite{hastings2006,foss-feig2015,else2020} and the vanishing of the characteristic dynamical time scale in the thermodynamic limit~\cite{cevolani2016}.
Conversely, algebraic interactions that decay fast enough
are effectively short range, and one recovers ballistic propagation of information~\cite{Chen+19,Tran+20,Kuwahara+20}.
The intermediate regime is, however, strongly debated,
and whether a form of causality emerges remains an open question.
Known bounds in principle allow for super-ballistic propagation~\cite{hastings2006,foss-feig2015,else2020} but they are challenged by numerical
simulations, which point towards a significantly slower propagation~\cite{hauke2013,schachenmayer2013,cevolani2015,schachenmayer2015b,buyskikh2016,luitzEmergentLocalitySystems2019, colmenarezLiebRobinsonBoundsOutoftime2020}.
The latter, however, reported different propagation scaling laws.
Microscopic mean-field theory suggests that these apparent contradictions may be attributed to the coexistence of several signals governed by different dynamical scaling laws~\cite{cevolani2018}.
This prediction, however, relies on a generic but non-universal form of the correlation functions and ignores beyond-mean-field effects.
Experiments performed with trapped ions have reported bounded propagation~\cite{richerme2014, jurcevic2014} but they are limited to very small systems, which prevents extraction of the scaling laws and closure of the debate.

The aim of this work is to characterize the spreading of quantum correlations in the intermediate regime of a long-range spin system numerically.
Specifically, we determine the scaling laws for the propagation  of a variety of observables in the long-range transverse Ising (LRTI) chain using matrix product state simulations.
We find that a weak form of causality emerges, characterized by generic algebraic scaling forms \(t \sim R^\beta\), where the specific value of the exponent \(\beta\) depends on the observable.
The spin-correlation and local-spin causal edges are  both sub-ballistic, with the same dynamical exponent (\(\betaCE=\betaSE > 1\)). In the vicinity of the edge, however, the local maxima propagate differently, \ie\ ballistically (\(\betam=1\)) for spin-spin correlations and super-ballistically (\(\betam<1\)) for local spins.
In contrast, the R\'enyi entanglement entropies always propagate ballistically, irrespective of the range of interactions, in both the local and quasi-local regime.
The analytic quasi-particle picture, based on linear spin wave theory (LSWT), accurately reproduces the numerics and provides a clear interpretation of the numerical results.
The different algebraic space-time patterns of correlation functions provide an unambiguous fingerprint of the dynamical regimes of the model, suggesting the emergence of a \textit{dynamical phase-diagram}. 
These correlation patterns can be directly measured  in state-of-the-art experiments.

\begin{figure}[t]
\includegraphics[width=0.9\linewidth]{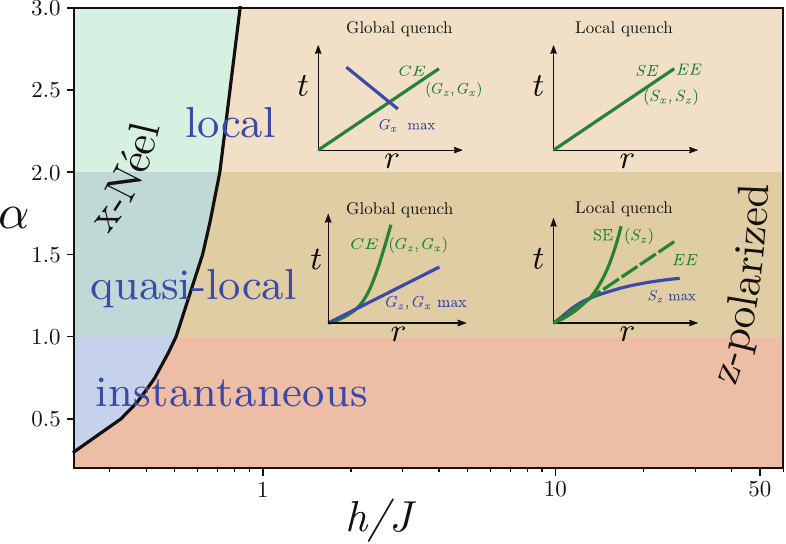}
\caption{\label{fig:phase_diagram}
Phase diagram and dynamical properties of the LRTI chain in the \((h/J,\alpha)\) plane. The \(x\)-N\'eel phase (shades of blue) is separated from the \(z\)-polarized phase (shades of orange) by a critical line in black. The three dynamical regimes are shaded differently.
As shown in the insets, the local and quasi-local regimes  are characterized by distinct algebraic scaling laws for the correlation and spin edges (CE and SE, respectively), their maxima in space-time (max), and the entanglement edge (EE), after global or local quench.
The different behaviors of these observables in different regions of the phase diagram provide clear, experimentally accessible, signatures of the local and quasi-local regimes.
}
\end{figure}

\lettersection{Model and approach}\label{sec:model}
The dynamics of the LRTI chain is governed by the Hamiltonian
\begin{equation}\label{H}
 \hat{H} = \sum_{R \neq R'} \frac{J}{|R-R'|^{\alpha}} \hat{S}^x_R \hat{S}^x_{R'} - 2h \sum_R \hat{S}^z_R,
\end{equation}
where \(\hat{S}_R^{j}\) (\(j=x,y,z\)) are the spin-\(1/2\) operators on lattice site \(R \in [0,N-1]\), \(N\)  is the system size,
\(J>0\) is the coupling energy, and \(h\) is the transverse field. 
It can be realized on various quantum simulation platforms,
including cold Rydberg gases~\cite{lukin2000,zoller2001,browaeys2016} and 
artificial ion crystals, where the exponent \(\alpha\) can be controlled via light-mediated interactions~\cite{deng2005,lanyon2011,schneider2012b,jurcevic2014,richerme2014}.
At equilibrium, the phase diagram of the LRTI chain comprises two gapped phases separated by a second order quantum phase transition, see Fig.~\ref{fig:phase_diagram} and Ref.~\cite{koffel2012}.
For low fields \(h\) and rather short-range couplings (high values of \(\alpha\)),
the nearest-neighbor anti-ferromagnetic couplings dominate
and the system forms a staggered N{\'e}el-ordered phase along the \(x\) direction.
For a large field \(h\) and long-range couplings (low values of \(\alpha\)), the spin-field interaction is favoured and a \(z\)-polarized phase is formed.
%
%
Out of equilibrium, the LSWT predicts three dynamical regimes (shaded by different colors in \cref{fig:phase_diagram})~\cite{hauke2013,cevolani2015}:
For \(\alpha \geq 2\) (local regime), the spin wave excitations are regular with bounded energies \(E_k\) and group velocities \(\Vg(k) = \partial_k E_k\) (\(k\) is the momentum). This regime is reminiscent of the short-range case where correlations spread at finite speed, giving rise to a linear causality cone~\cite{lieb1972,hastings2006}.
For \(\alpha<1\) (instantaneous regime), \(E_k\) features an algebraic, infrared divergence. There is no characteristic
time scale and correlations spread arbitrarily fast.
Finally, for \(1 \leq \alpha<2\) (quasi-local regime), \(E_k\) is bounded but \(\Vg(k) \) diverges as \(\Vg(k) \sim 1/k^{2-\alpha}\).
Whether some form of causality emerges in the quasi-local regime remains debated~\cite{hauke2013,jurcevic2014,richerme2014,cevolani2015,
maghrebi2016,cevolani2016,buyskikh2016},
and, in the following, we mainly focus on this case.

Except where otherwise indicated, the results discussed below are obtained using time-dependent variational principle (TDVP) simulations within a matrix-product state (MPS) framework~\cite{haegeman2012,koffel2012,hauke2013}. Convergence of the calculations with the bond dimension has been systematically checked.
Our results are summarized on the dynamical phase diagram of Fig.~\ref{fig:phase_diagram}.
The main features of the observable that we consider are all described by an algebraic space-time dependence of their edges and maxima, shown in the insets. 

\begin{figure}[t]
\includegraphics[width = \columnwidth]{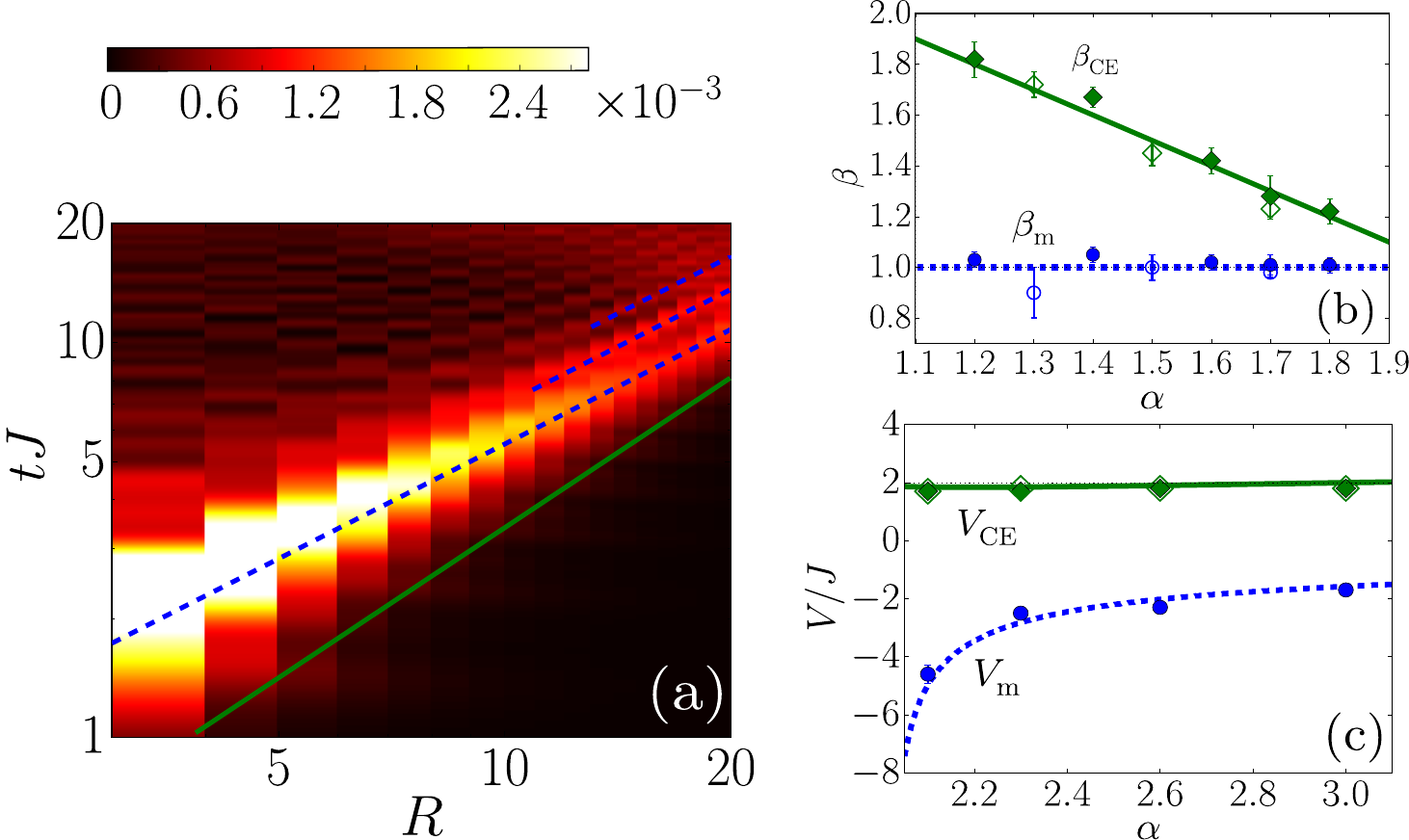}
\caption{\label{fig:SpinCorr} 
Spreading of spin correlations in a global quench, with system size \(N=48\).
(a)~TDVP results for \(G_z\)
and a quench from \((h/J)_\mathrm{i} = 50\) to \((h/J)_{\mathrm{f}} = 1\)
in the quasi-local regime at \(\alpha = 1.7\).
The solid green and dashed blue lines are fits to the CE and the extrema, respectively.
(b)~Dynamical exponents \(\betaCE\) (green diamonds) and \(\betam\) (blue disks), fitted
form results as in panel~(a) for \(G_z\) (empty symbol) and \(G_x\) (full symbol), 
and comparison to the LSWT predictions (solid green and dashed blue lines).
(c)~Spreading velocities \(\VCE\) (green diamonds) and \(\Vm\) (blue disks),
in the local regime and comparison to the LSWT predictions (solid green and dashed blue lines).
}
\end{figure}

\lettersection{Spin correlations and global quench}
We first consider the spreading of the connected spin correlations,
\(G_j(R,t) =  G_j^0(R,t) - G_j^0(R,0)\), with \(G_j^0(R,t) = \langle \hat{S}^j_R(t)\hat{S}^j_0(t)\rangle - \langle \hat{S}^j_R(t) \rangle \langle \hat{S}^j_0(t) \rangle\), along the directions \(j=x,z\).
Figure~\ref{fig:SpinCorr}(a) shows a typical TDVP result for \(G_z(R,t)\),
for a global quench in the quasi-local regime, \(\alpha=1.7\), from \((h/J)_\textrm{i}=50\) to \((h/J)_\textrm{f}=1\), both in the \(z\)-polarized phase. The initial state of the system is the ground state of the Hamiltonian with \((h/J)_\textrm{i}\).
Similar results are found when scanning the values of \((h/J)_\textrm{i}\), \((h/J)_{\textrm{f}}\), and \(1<\alpha<2\).
The correlation pattern shows a series of maxima which propagate algebraically [notice the log-log scale in Fig.~\ref{fig:SpinCorr}(a)].
In the asymptotic long-time and long-distance limits, they are well fitted by the scaling law
\(t \sim R^{\betam}\) (dashed blue lines).
The correlation edge (CE), which sets the causality horizon, cannot, however, be inferred from the behavior of the maxima~\cite{despres2019}.
To find the CE, we track the points of the \(R-t\) plane where the correlations reach a fraction \(\epsilon\) of their maximum.
Scanning \(\epsilon \in [0.01, 0.1]\), we find that the CE is well fitted by the algebraic scaling law \(t \sim R^{\betaCE}\) (solid line), with \(\betaCE\) nearly independent of \(\epsilon\)~\cite{note:SupplMat}.
Figure~\ref{fig:SpinCorr}(b) shows the results of the fits (empty symbols) versus the exponent \(\alpha\).

The numerical results for \(G_x(R,t)\) have an extra checkerboard structure inside the causal region due to the antiferromagnetic coupling in Eq.~(\ref{H}).
Once this structure is taken into account, we can identify a CE and local maxima, and extract the corresponding exponents. The results are similar to those found for \(G_z(R,t)\), see filled symbols in Fig.~\ref{fig:SpinCorr}(b)~\cite{note:SupplMat}.

Comparing the fitted dynamical exponents \(\betam\) and \(\betaCE\) to the predictions of the LSWT~\cite{cevolani2018}, we find an excellent agreement, as shown in Fig.~\ref{fig:SpinCorr}(b).
While the maxima spread ballistically, \(\betam \simeq 1\), the CE is sub-ballistic with \(\betaCE \simeq \betaSWT = 3-\alpha>1\).
This is characteristic of gapped long-range models, such as the LRTI model in the z-polarized phase~\cite{cevolani2018}.
It confirms the emergence of a weak, slower-than-ballistic form of causality for \(1<\alpha<2\).

In the local regime, \(\alpha \geq 2\), we find that both the maxima and the CE spread ballistically, \(\betam \simeq \betaCE \simeq 1\). The spreading velocities are, however, different from each other,
which is characteristic of a non-phononic excitation spectrum.
The CE velocity is \(\VCE \simeq 2V_{\mathrm{g}}(k^*)\), \ie\ twice the maximum group velocity of the spin waves, while that of the maxima is \(\Vm \simeq 2V_\varphi(k^*)\), \ie\ twice the phase velocity \(V_{\varphi}(k) = E_k/k\) at the quasi-momentum \(k^*\) where the group velocity is maximum, see Fig.~\ref{fig:SpinCorr}(c).

We have also performed calculations for large quenches.
While the LSWT is well justified only for weak quenches, we have found that
the dynamical exponents it predicts are extremely robust, even for quenches across the critical line.
We found dynamical scaling laws in very good agreement with those reported on Figs.~\ref{fig:SpinCorr}(b) and (c), although the signal is blurred as compared to weak quenches owing to the proliferation of quasi-particles when the quench amplitude increases~\cite{note:SupplMat}.

\begin{figure}[t!]
\includegraphics[width = \columnwidth]{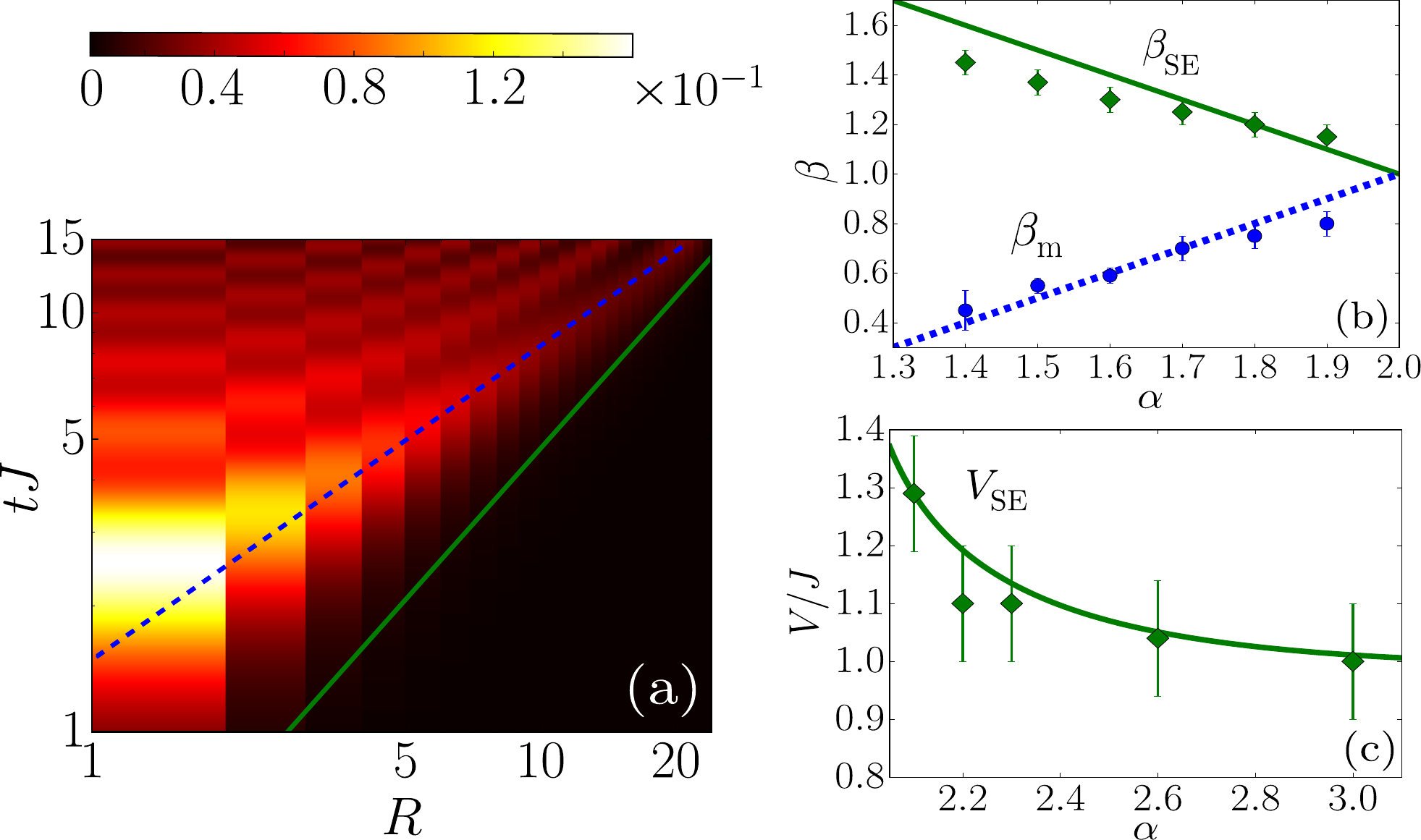}
\caption{\label{fig:numerics2}
Spreading of the local magnetization for a local quench, with system size \(N=48\).
(a)~TDVP results for the quantity \(1/2 - \langle \hat{S}^z_{R}(t) \rangle\)
versus the time and the distance from the flipped spin for \(h/J=50\) and \(\alpha = 1.8\). 
(b)~Dynamical exponents \(\betaSE\) (green diamonds) and \(\betam\) (blue disks), fitted
form results as in panel~(a) and comparison to the LSWT predictions (solid green and dashed blue lines).
(c)~Spreading velocity \(\VSE\) (green diamonds) and comparison to the LSWT prediction in the local regime (solid green line).
}
\end{figure}

\lettersection{Local magnetization and local quench}
We now consider the dynamics of another quantity, namely the local magnetization \(\langle \hat{S}^z_{R}(t) \rangle\), and perform a local quench.
We initialize the system in the ground state of the \(z\)-polarized phase and flip the central spin.
Figure~\ref{fig:numerics2}(a) shows a typical TDVP result for the quantity
\(1/2-\langle \hat{S}^z_{R}(t) \rangle\)
versus the time \(t\) and the distance \(R\) from the flipped spin, in the quasi-local regime.
The result displays, as in the case of global quenches, a twofold algebraic structure.
Fitting the maxima (dashed blue line) and the spin edge (SE, solid green line) as previously,
we extract the dynamical exponents \(\betam\) and \(\betaSE\) plotted in Fig.~\ref{fig:numerics2}(b) (blue disks and green diamonds, respectively).
The SE follows the same sub-ballistic scaling law as the CE in the previous case,
\(\betaSE \simeq 3-\alpha\) (solid green line).
In contrast, the maxima of the local spin spreads faster than those of the spin correlations,
and here we find \(\betam < 1\), corresponding to a super-ballistic propagation.

To understand this behavior, let us first recall that the standard LSWT generically predicts a super-ballistic propagation of the maxima in gapless systems~\cite{cevolani2018}. Here, although the LRTI model is gapped, the initial state \(\ket{\Psi_0} \simeq \ket{\uparrow ... \uparrow \, \downarrow \, \uparrow ... \uparrow}\) of the local quench is orthogonal to the ground state \(\ket{\Psi_\textrm{\tiny GS}} \simeq \ket{\uparrow \, \uparrow ... \uparrow}\), and thus lives in the first excited manifold.
The ground state and the gap are thus irrelevant, and we may expect super-ballistic propagation of the maxima 
consistent with the TDVP results.
This argument applies to any observable for such a local quench.
More specifically, the LSWT applied to \(\langle \hat{S}^z_{R}(t) \rangle\) for the local quench yields \(\betam = \alpha-1\), in very good agreement with the TDVP results, see Fig.~\ref{fig:numerics2}(b)~\cite{note:SupplMat}.

In the local regime (\(\alpha > 2\)) we find that \(\langle \hat{S}^z_{\tilde{R}}(t) \rangle\) propagates ballistically.
Both this property and the SE velocity extracted from the TDVP calculations are in good agreement with  the LSWT analysis, see Fig.~\ref{fig:numerics2}(c).
Note that in this regime, we do not observe maxima propagating at a different velocity. This is also consistent with the LSWT analysis.
In the local regime, the quantity \(\langle \hat{S}^z_{R}(t) \rangle\) is the sum of several contributions, each with a twofold structure but that are in phase quadrature and cancel each other.
A similar effect has been found for density correlations deep in the Mott insulator phase
of the Bose-Hubbard model~\cite{despres2019,note:SupplMat}.

\begin{figure}[t]
\includegraphics[width = \columnwidth]{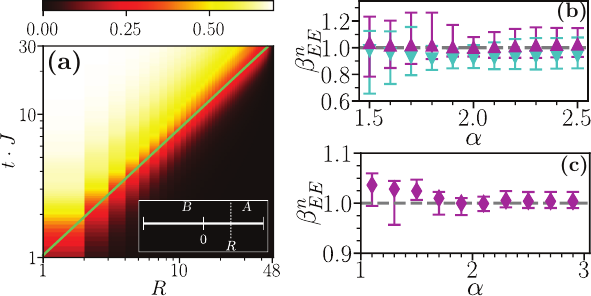}
\caption{\label{fig:numerics3}
  Spreading of the von Neumann entanglement entropy \(\mathcal{S}_{n=1}(R,t)\) for the same quench as in Fig.~\ref{fig:numerics2}(a).
  (a)~TDVP results for 
  \(h/J=50\), \(\alpha=1.8\), and a system size \(N=96\).
  The solid green line marks a power law fit to the EE with \(\epsilon = 0.5\), yielding
  \(\beta_{EE}^n = 0.899 \pm 0.005\).
  (b)~Dynamical exponents for the EE obtain via TDVP (cyan downwards triangles) and LSWT (magenta upwards triangles), with error bars corresponding to the variations with respect to the threshold \(\epsilon\).
  (c)~Dynamical exponents obtained via LSWT for a larger system, \(N=512\).
}
\end{figure}

\lettersection{Entanglement entropy}
We finally study the spreading of quantum information after the same local quench. It may be measured via the R\'enyi entropies of the reduced density matrix of a block,
\begin{equation}\label{eq:Renyi}
 \mathcal{S}_n(R,t) = \frac{1}{1-n} \log\!\big\{\tr[ \hat{\rho}^n(R,t) ]\big\} \,,
\end{equation}
with 
\(n > 0\) and \(\hat{\rho}(R,t) = \hat{\rho}_A = \tr_{B}(\ketbra{\Psi(t)}) \)
the reduced density matrix at time \(t\) of the subsystem \(A = [R, R+1,\ldots,N/2]\).
The position \(R\) is measured from the flipped central spin and \(B\) denotes the complementary subsystem, see Inset in Fig.~\ref{fig:numerics3}(a).

Figure~\ref{fig:numerics3}(a) shows a typical TDVP result for the von Neumann entropy, \(n \rightarrow 1\) \footnote{The von Neumann entropy also reads as \(\mathcal{S}_{\textrm{vN}}=\mathcal{S}_{n \rightarrow 1}(R,t) = -\mathrm{Tr}\left \{ \hat{\rho}(R,t) \mathrm{log}\left[ \hat{\rho}(R,t) \right] \right \}\)}, and the same local quench as in Fig.~\ref{fig:numerics2}(a). 
Similar results are found for the other R\'enyi entropies~\cite{note:SupplMat}.
We find that \(\mathcal {S}_{n}(R,t)\) is a monotonic function of both position and time, and no local maxima such as those found previously for the correlation function and the magnetization are observed. This is consistent with a causal spreading of information and the expectation that the entanglement entropy decreases with the size of the smaller partition (\(A\)).
The entanglement edge (EE) is clearly visible in Fig.~\ref{fig:numerics3}(a) and a fit to  the algebraic scaling law \(t \sim R^{\betaEE{n}}\) allows us to extract the exponent.
Varying the threshold in a wide range, \(\epsilon \in [0.2, 0.8]\),
we find \(\betaEE{n} \approx 1\) within error bars, irrespectively of the interaction range, see cyan points in Fig.~\ref{fig:numerics3}(b).
The error bars correspond to the variation of \(\betaEE{n}\) with \(\epsilon\), which is due to finite-size effects (see below).
Our results imply that the propagation of entanglement is close to ballistic in both the local and quasi-local regimes.
This contrasts with the behavior of the one- and two-point observables considered thus far, which exhibit sub-ballistic causal edges.
It is a direct consequence of the fact that the bipartite entanglement entropy is a highly non-local quantity, which takes into account all entangled pairs on either side of the bipartition \(R\)~\cite{calabrese2006,alba_entanglement_2017,bastianello_spreading_2018, calabrese_entanglement_2020}.

More precisely, we have analytically computed the entanglement entropy within LSWT~\cite{note:SupplMat}.
The results are in good agreement with those extracted via TDVP for the same system sizes, see Fig.~\ref{fig:numerics3}(b).
The ballistic spreading of entanglement is further confirmed by the LSWT results obtained in much larger systems, with significantly smaller error bars, see Fig.~\ref{fig:numerics3}(c).
Bipartitioning the quenched initial state within LSWT yields an entanglement Schmidt rank bounded by \(2\). This is consistent with our TDVP results at all times
and the saturation of the entanglement entropies to \(\mathcal{S}_n(R,t \rightarrow\infty) \simeq \log(2) \simeq 0.69\), see Fig.~\ref{fig:numerics3}(c).
The two eigenvalues of the reduced density matrix \(\hat{\rho}_A\), \(\lambda_1(R,t)\) and \(\lambda_2(R,t)=1-\lambda_1(R,t)\), can then be computed analytically.
In the asymptotic limit and for not too small values of \(R/t\), 
we find \(\lambda_2(R,t) \propto t^{\frac{1}{2-\alpha}}\zeta(\frac{3-\alpha}{2-\alpha},R)\sim (t/R)^{\frac{1}{2-\alpha}}\), with \(\zeta\) the Hurwitz zeta function~\cite{Hurwitz1932,Magnus1966}.
Hence, the \(n\)-order R\'enyi entropy is a function of the ratio \(R/t\),
\begin{equation}\label{eq:RenyiEnd}
\mathcal{S}_{n}(R,t) \simeq \frac{1}{1-n} \log\!\big\{ \lambda_1\!\left({R}/{t}\right)^n + \lambda_2\!\left({R}/{t}\right)^n \big\} \, .
\end{equation}
This confirms the ballistic propagation of the entanglement entropy
(\(\betaEE{n}=1\)) consistent with the results of Fig.~\ref{fig:numerics3} for \(n=1\)
and other R\'enyi orders \(n\)~\cite{note:SupplMat}.

\lettersection{Conclusion}
Our results show the emergence of a weak form of causality in the intermediate regime of the long-range Ising model, characterized by algebraic propagation laws with exponents that depend on the observables and the range of interactions. While local spins and spin correlations both have a sub-ballistic propagation edge, \(t \propto R^\beta\), with \(\beta >1\), the causal region is characterized by local maxima propagating super-ballistically and ballistically, respectively.
The distinction between the causal edge and the local maxima, which can show drastically different dynamical behaviors, is thus pivotal in the characterization of causality in long-range quantum systems.
In contrast, the propagation of  entanglement is ballistic in both the local and quasi-local regimes, and the causal region is featureless. 

These results call for future experimental and theoretical work.
On the one hand, our predictions are directly relevant to quantum simulators using for instance trapped ions, where the interaction range can be controlled. While analysis of spin and correlation spreading in first experiments have been limited by finite-size effects~\cite{jurcevic2014,richerme2014}, systems with more than 50 ions, comparable to the system size used in our simulations, are now accessible~\cite{zhang2017} and R\'enyi entanglement entropies can now be measured in trapped ion platforms~\cite{linke2018,brydges2019} and digital quantum computers~\cite{vovroshConfinementEntanglementDynamics2020}.
On the other hand, it would be interesting to further test the robustness of the observed algebraic scaling laws by quantitatively investigating the dependence (if any) of the exponents on the strength of the quenches, the phases of the model and their values in different models, such as the long-range XY~\cite{schachenmayer2013,cevolani2018}, Heisenberg~\cite{yangDeconfinedSpinonsCoherent2021,yangTopologicalMagneticallyOrdered2021}, and Hubbard~\cite{manmana2009,cevolani2015} models, as well as in dimensions higher than one.
This extended analysis could allow the identification of \textit{dynamical universality classes}, \ie\ models which share the same algebraic laws for correlations out-of-equilibrium.

\acknowledgments

We thank the CPHT computer team for valuable support. The numerical calculations were performed using HPC resources from GENCI-CINES (Grant 2019-A0070510300). We acknowledge use of the {ITensors.jl} software package~\cite{Fishman+20}.
L.T.~acknowledges support from the Ramón y Cajal program RYC-2016-20594, the ``Plan Nacional Generación de Conocimiento'' PGC2018-095862-B-C22 and the State Agency for Research of the Spanish Ministry of Science and Innovation through the ``Unit of Excellence María de Maeztu 2020-2023'' award to the Institute of Cosmos Sciences (CEX2019-000918-M).

J.T.S.~and J.D.~contributed equally to this work.

\bibliographystyle{revtexlsp}
\bibliography{refs,notes}

\appendix
  
 \renewcommand{\theequation}{S\arabic{equation}}
 \setcounter{equation}{0}
 \renewcommand{\thefigure}{S\arabic{figure}}
 \setcounter{figure}{0}
 \renewcommand{\thesection}{S\arabic{section}}
 \setcounter{section}{0}
 \onecolumngrid
 \newpage

\begin{center}
{\large \textbf{Supplemental Material for ``Spreading of Correlations and Entanglement in the Long-Range Transverse Ising Chain''} }
\end{center}
\vspace*{1.cm}

In this Supplemental Material, we provide additional technical details of the methodology and analysis underlying the central results discussed in the main paper.
In Sec.~\ref{sec:edge}, we detail the fitting procedure to extract the scaling exponent of the spin-correlation, local magnetization, and entanglement edges.
In Sec.~\ref{gx_qlr_lr_regime}, we discuss the spreading of spin correlations in the \(x\) direction and in Sec.~\ref{large_quenches} results for large quenches.
Sections~\ref{local_mag} and \ref{app:density_matrix_local} outline the LSWT theory for the spreading of the local magnetization \(\langle S^z_R(t)\rangle\) and entanglement entropies \(\mathcal{S}_{n}(R,t)\), respectively, after a local quench.
Finally, Sec.~\ref{sec:ent_entropies} gives additional TDVP and LSWT results for the spreading of entanglement considering various R\'enyi entropies.

\section{Determination of the edge}
\label{sec:edge}
\begin{figure}[h!]
\includegraphics[scale = 0.4]{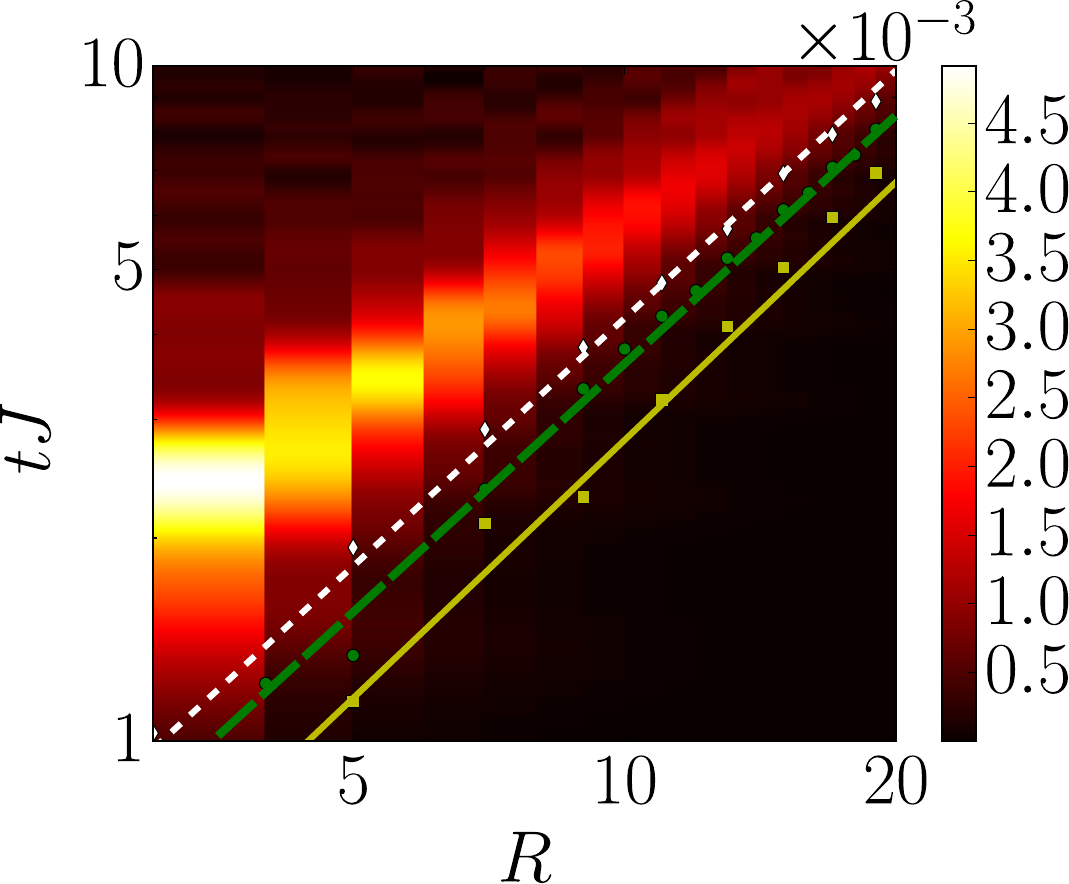}
\caption{
Spreading of the \(G_z\) spin correlation function in the quasi-local regime (\(\alpha = 1.7 < 2\)),
for a global quench from \((h/J)_\mathrm{i} = 50\) to \((h/J)_{\mathrm{f}} = 1\), both in the \(z\) polarized phase [same data as in the Fig.~\ref{fig:SpinCorr}(a) of the main paper].
The squares, disks, and diamonds indicate points where \(G_z(R,t)\) reaches a fraction \(\epsilon\) of its maximum value for various values of \(\epsilon\).
The corresponding lines are linear fits to these points in log-log scale, consistently with Eq.~(\ref{eq:algebraicfit}).
The brown squares correspond to \(\epsilon = 2.6\%\) for which we find \(\betaCE \simeq 1.27\),
the green disks to \(\epsilon = 6.5\%\) for which we find \(\betaCE \simeq 1.23\),
and the white diamonds to \(\epsilon = 11.2\%\)  for which we find \(\betaCE \simeq 1.22\).
\label{gz_qlr_eps}
}
\end{figure}

For all the data reported in the main paper, we have determined the spin-correlation,  local magnetization, and entanglement edges by tracking the ensemble of points in the \(R\)--\(t\) plane where the signal reaches a fraction \(\epsilon\) of its maximal value.
Since this threshold line depends on the value of \(\epsilon\), we have systematically scanned \(\epsilon\), e.g.\ from \(0.01\) to \(0.12\) for \(G_z(R,t)\) and  from \(0.20\) to \(0.80\) for \( \mathcal{S}_n(R,t)\).
In all the cases considered in this work, we find that the edge is well fitted by the algebraic law
\begin{equation}\label{eq:algebraicfit}
t =  a \times R^{\beta}.
\end{equation}
While the coefficient \(a\) fundamentally depends on \(\epsilon\), the scaling law exponent \(\beta\) is nearly independent of \(\epsilon\).
An example of such an analysis, plotted in log-log scale, is shown in Fig.~\ref{gz_qlr_eps} where the various lines correspond to different values of \(\epsilon\). The fact that they are parallel straight lines validates the scaling law \(t \sim \times R^{\betaCE}\) with an exponent \(\betaCE\) nearly independent of \(\epsilon\).

\section{Spreading of the \(\mathbf{G_x}\) spin correlation function} 
\label{gx_qlr_lr_regime}

In this section, we discuss analytic and numerical results for the spreading of spin correlations along the \(x\) direction,
\begin{equation}
G_x(R,t) =  G_{x}^0(R,t) - G_{x}^0(R,0)
\qquad \textrm{with} \qquad
G_{x}^0(R,t) = \langle \hat{S}^x_R(t)\hat{S}^x_0(t)\rangle - \langle \hat{S}^x_R(t) \rangle \langle \hat{S}^x_0(t) \rangle,
\end{equation}
where \(\hat{S}^x_R(t)\) is the spin operator along \(x\) at position \(R\) and time \(t\).

\subsection{Linear spin wave theory}\label{LSWT}
In the \(z\)-polarized phase, the LRTI model can be diagonalized using a Holstein-Primakoff transformation~\cite{holstein1940}, 
\begin{equation}
\hat{S}^x_R \simeq \frac{1}{2} \left( \hat{a}^{\dag}_R + \hat{a}_R \right)
\qquad , \qquad
\hat{S}^y_R \simeq \frac{-1}{2i} \left( \hat{a}^{\dag}_R - \hat{a}_R \right)
\qquad , \quad \textrm{and} \qquad
\hat{S}^z_R \simeq \frac{1}{2} - \hat{a}^{\dag}_R \hat{a}_R,
\label{HP}
\end{equation}
where \(\hat{a}_R(t)\) is the annihilation operator of a boson at site \(R\) and time \(t\).
Inserting these formulas into the LRTI Hamiltonian [Eq.~(\ref{H}) of the main paper], we find a quadratic Bose Hamiltonian, see for instance Refs.~\cite{hauke2013,cevolani2016} for details.
The latter is readily diagonalized using a standard Bogoliubov transformation~\cite{bogoliubov1947}.
The elementary excitations are magnons (spin-wave excitations), characterized by the gapped spectrum
\begin{equation}
 E_k = 2\sqrt{h[h+JP_{\alpha}(k)]} = \sqrt{\mathcal{A}_k^2 - \mathcal{B}_k^2}
 \qquad , \quad \mathrm{with} \qquad
 \mathcal{A}_k = J P_{\alpha}(k) + 2h
 \qquad \mathrm{and} \qquad
 \mathcal{B}_k = J P_{\alpha}(k),
 \label{exc_spec_z}
\end{equation}
where \(P_{\alpha}(k) = \sum_R e^{ikR}/R^\alpha\) denotes the Fourier transform of the long-range term in the Hamiltonian.

For a quench on the exchange amplitude \(J\) from \(J_{\mathrm{i}}\) to \(J_{\mathrm{f}}\),
the \(G_x(R,t)\) correlation function can be cast in the form~\cite{cevolani2018}
\begin{equation}
 G_x(R,t) \simeq g(R) -\int_{-\pi}^{\pi} \frac{\mathrm{d}k}{2\pi} \mathcal{F}(k)
 \left\{ \frac{e^{i(kR+2E_k^{\mathrm{f}}t)} + e^{i(kR-2E_k^{\mathrm{f}}t)}}{2} \right\}
 \quad , \quad \mathrm{with} \quad
 \mathcal{F}(k) = \frac{h(J_\mathrm{i}-J_\mathrm{f})P_{\alpha}(k)}{8[h+J_\mathrm{f}P_{\alpha}(k)]
\sqrt{h[h + J_{\mathrm{i}}P_{\alpha}(k)]}},
 \label{gx_analy}
\end{equation}
where the index "i" refers to the pre-quench (initial) Hamiltonian and the index "f" to the post-quench (final) Hamiltonian.
The function \(\mathcal{F}(k)\) gives the weight of each quasi-particle according to their mode \(k\).
It depends on the observable and the quench.

To characterize the asymptotic behavior, \(R,t \rightarrow +\infty\) along a constant line \(R/t\), one can then rely on the stationary-phase approximation, applied to Eq.~\eqref{gx_analy}.
It yields
\begin{equation}
 G_x(R,t) \sim \frac{\mathcal{F}(k_{\mathrm{sp}})}{(|\partial^2_k E_{k_{\mathrm{sp}}}|t)^{1/2}} \cos(k_{\mathrm{sp}}R
 - 2E_{k_{\mathrm{sp}}}t + \phi),
 \label{gx_spa}
\end{equation}
where \(\phi\) is a constant phase irrelevant to our study,
\(k_{\mathrm{sp}}\) is the stationary-phase quasi-momentum,
given by the solution of the equation \(2V_{\mathrm{g}}(k_{\mathrm{sp}}) = 2\partial_{k} E_{k_{\mathrm{sp}}} = R/t > 0\) with \(\Vg\) the group velocity associated to the spin-wave excitations.

\bigskip
\paragraph{Quasi-local regime~--~}
\label{SM:QLregAnalytics}
The quasi-local regime corresponds to the case where the quasi-particle energy \(E_k\) is finite over the whole first Brillouin zone, \(k \in [-\pi,+\pi]\), but the group velocity \(\Vg(k)\) presents a divergence.
The space-time behavior of \(G_x\) in the vicinity of the CE (correlation edge)
is dominated by the quasi-particles propagating with the highest group velocities.
Due to the infrared divergence, only the limit behavior in \(k\rightarrow 0\) is relevant.
There the long-range term reads as
\(P_{\alpha}(k) \approx P_{\alpha}(0) + P_{\alpha}' |k|^{\alpha-1}\),
with \(P_{\alpha}(0)>0\), \(P'_{\alpha} <0\),
and the excitation energy is \(E_k \simeq \Delta - c |k|^z\),
where \(\Delta = 2\sqrt{h[h+J  P_{\alpha}(0)]} > 0\) is the gap,
\(z =  \alpha-1 \geq 0\) is the dynamical exponent,
and \(c = \sqrt{h/[h+JP_{\alpha}(0)]}J|P_{\alpha}'|>0\) is a prefactor.
Then, computing \(k_{\mathrm{sp}}\) and injecting it into Eq.~\eqref{gx_spa},
we find
\begin{equation}
G_x(R,t) \sim \frac{t^{\gamma}}{R^{\chi}} \cos \Big[ A_z \Big( \frac{t}{R^z} \Big)^{\frac{1}{1-z}}  - 2 \Delta t + \phi \Big],
\label{infrared}
\end{equation}
with
\begin{equation}
A_z = 2c (2cz)^{z/(1-z)}(1-z)
 \qquad \mathrm{,} \qquad\gamma = \frac{\nu +1/2}{1-z}
 \qquad \mathrm{,} \quad \mathrm{and} \qquad
 \chi = \frac{\nu + (2-z)/2}{1-z},
\end{equation}
where \(\nu \geq 0\) is the scaling exponent of the amplitude function in the infrared limit,
\(\mathcal{F}(k) \sim |k|^{\nu}\).
It follows from Eq.~\eqref{gx_analy} and the approximation of \(P_{\alpha}\) in the infrared limit
that \(\nu = 0\) for the \(G_x\) spin correlation function.
On the one hand, the CE is found by imposing the condition that the prefactor is constant.
The latter leads to the algebraic form \(t \propto R^{\beta_{\mathrm{CE}}}\) with \(\betaCE = \chi/\gamma = 3-\alpha\).
Since \(1 \leq \alpha < 2\) in the quasi-local,
the CE is always sub-ballistic, \ie\ \(\beta_{\mathrm{CE}} > 1\).
On the other hand, the spreading law of the local extrema is determined by the equation
\begin{equation}
A_z \Big( \frac{t}{R^z}\Big)^{\frac{1}{1-z}}  - 2 \Delta t + \phi = \mathrm{cst}
\qquad \mathrm{leading~to} \qquad
A_z \Big( \frac{t}{R} \Big)^{\frac{z}{1-z}}  - 2\Delta \rightarrow 0.
\label{cond_maxima}
\end{equation}
The maxima are thus ballistic, \ie\ \(t \sim R^\betam\) with \(\betam=1\).
\bigskip
\paragraph{Local regime~--~}
\label{SM:LocRegAnalytics}
In the local regime, corresponding to the case where both the quasi-particle energy \(E_k\) and the group velocity \(\Vg(k)\) are finite over the whole first Brillouin zone,
there exists a quasi-momentum \(k^*\) such that the group velocity is maximum,
\(\Vg(k^*) = \mathrm{max}_k[\Vg(k)]\). 
Hence, the stationary-phase condition \(2\Vg(k_{\mathrm{sp}}) = R/t\) has a solution only for \(R/t \leq 2\Vg(k^*)\).
The CE is determined by the spreading of the quasi-particles with a quasi-momentum \(k_{\mathrm{sp}} \simeq k^*\).
It is thus ballistic, with the CE velocity \(\VCE = 2\Vg(k^*)\).
Moreover, in the vicinity of the CE, the motion of the local maxima is determined by the phase factor in Eq.~\eqref{gx_spa} with \(k_{\mathrm{sp}} = k^*\).
It follows that the local extrema propagate ballistically at the velocity \(\Vm = 2V_{\varphi}(k^*)\)
where \(V_{\varphi}(k) = E_k/k\) is the quasi-particle phase velocity.
Note that here, \(k^*<0\) and \(\Vm = 2V_{\varphi}(k^*) = 2E_{k^*}/k^* < 0\).

\subsection{Numerical TDVP results}
\label{SM:Gx.numerics}
We have performed numerical TDVP calculations for the spreading of the spin correlation function along \(x\) for the same parameters as for its counterpart in the \(z\) direction, discussed in the text.

\begin{figure}[h!]
\includegraphics[scale = 0.35]{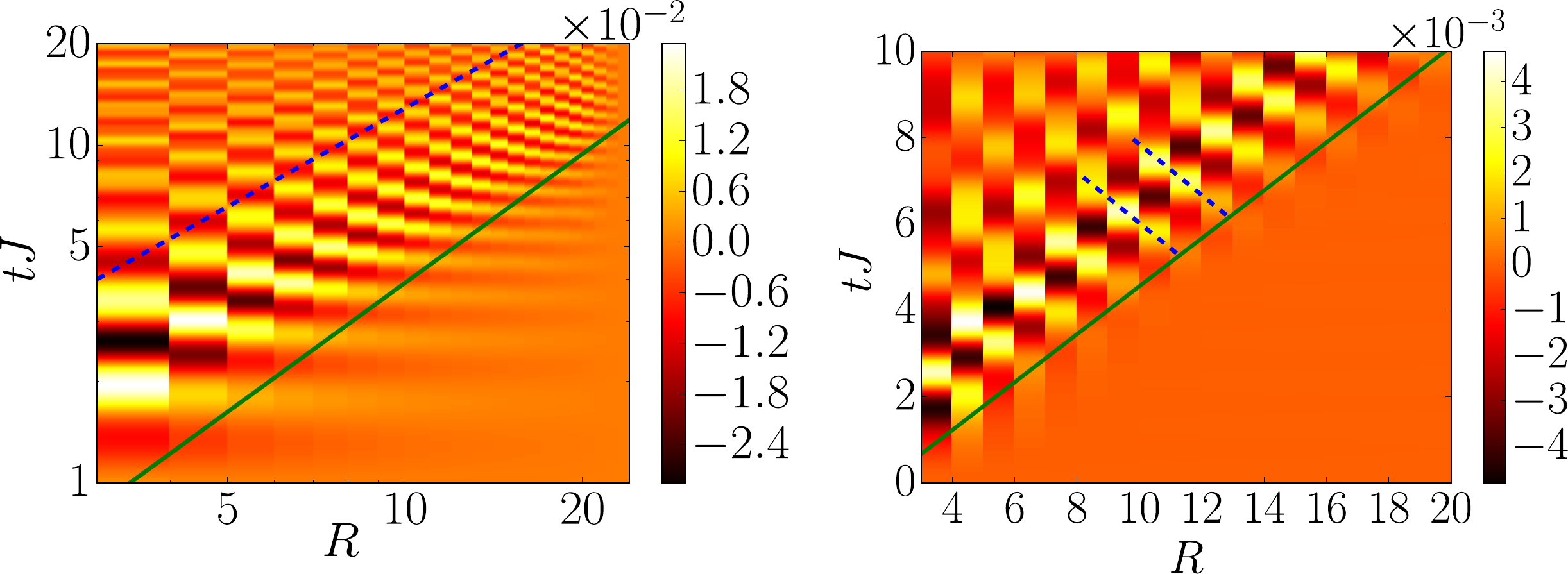}
\caption{
Spreading of the \(G_x\) spin correlations.
Left panel (log-log scale): Quasi-local regime with \(\alpha = 1.7 < 2\) for a global quench from \((h/J)_\mathrm{i} = 50\) to \((h/J)_{\mathrm{f}} = 1\), both in the \(z\)
polarized phase of the LRTI model.
Right panel (linear scale): Local regime with \(\alpha = 3 < 2\)
for a global quench starting from \((h/J)_\mathrm{i} = 0.9\) to \((h/J)_{\mathrm{f}} = 1\).
The solid green and dashed blue lines are algebraic (for \(\alpha<2\)) or linear (for \(\alpha>2\)) fits to the CE and the extrema, respectively.
\label{fig_gx}
}
\end{figure}

The left panel of Fig.~\ref{fig_gx} shows a typical result, plotted in log-log scale, in the quasi-local regime. The underlying checkerboard-like structure is characteristic of the \(G_x\) function and was similarly found using meanfield calculations in Ref.~\cite{cevolani2018}.
In spite of this complex structure, we can identify a CE, using the same \(\epsilon\) method used for the \(G_z\) correlation function, as well as propagating local maxima.
The green solid and dashed blue lines correspond to alegbraic fits to the CE and the local extrema, respectively. The associated scaling law exponents \(\betaCE\) and \(\betam\) are in very good agreement with the theoretical ones discussed in Sec.~\ref{SM:QLregAnalytics}, see filled symbols in Fig.~\ref{fig:SpinCorr}(b) of the main paper.

The right panel of Fig.~\ref{fig_gx} shows a typical result in the local regime, now plotted in linear scale.
The solid green and dashed blue lines are linear fits to the CE and local extrema, respectively.
The corresponding numerical velocities \(\VCE\) and \(\Vm\) are in good agreement with the theoretical ones, see filled symbols in Fig.~\ref{fig:SpinCorr}(c) of the main paper.

\section{Large quenches}
\label{large_quenches}

In this section, we present numerical results for large quenches using the same TDVP calculations as for the results of the main paper.

\begin{figure}[!ht]
\centering
\includegraphics[width = \columnwidth]{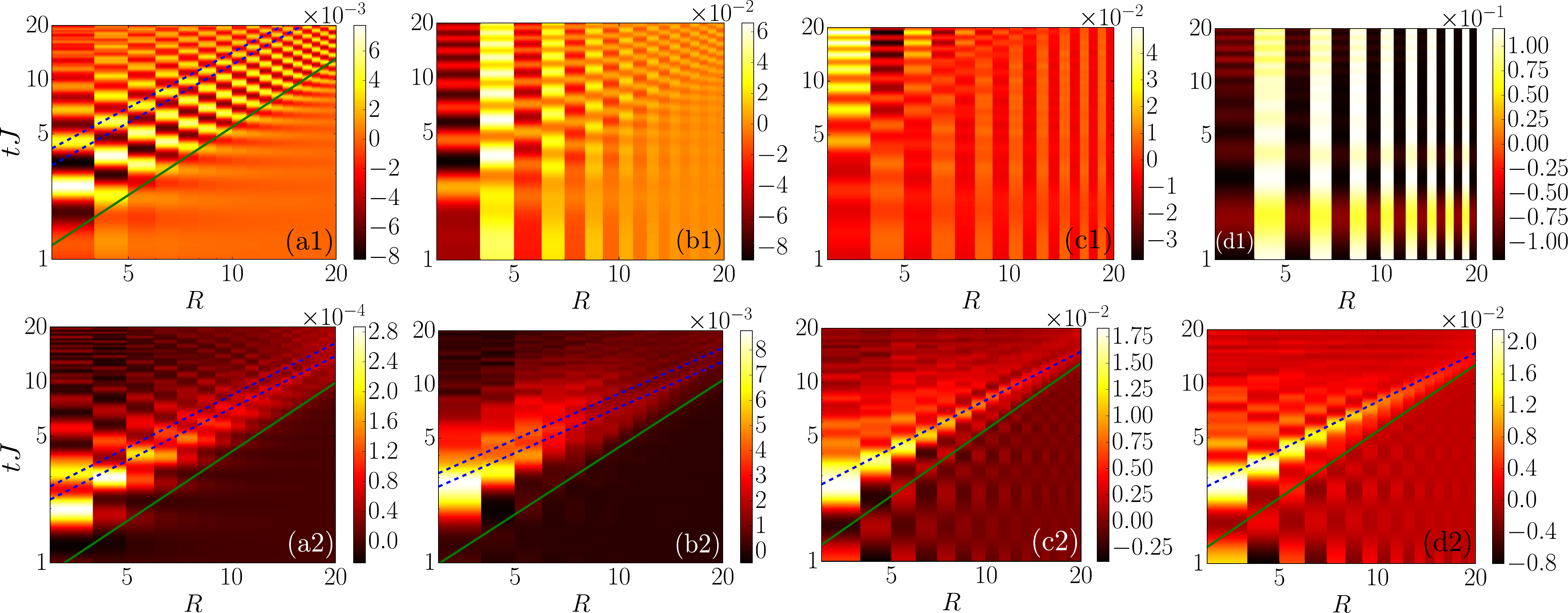}
\caption{
  Spreading of the \(G_x\) (upper row) and \(G_z\) (lower row) spin correlations in the quasi-local regime, \(\alpha = 1.7\) (log-log scale).
  The quench is performed from an increasing value of the initial parameter \((h/J)_{\mathrm{i}}\), ranging from the \(z\) polarized phase to the \(x\)-N\'eel phase,
  to the fixed value \((h/J)_{\mathrm{f}} = 1\), corresponding to the \(z\) polarized phase.
  (a)~\(\hat{H}_{\mathrm{i}}\) deep in the \(z\) polarized phase, \((h/J)_{\mathrm{i}} = 0.8\);
  (b)~\(\hat{H}_{\mathrm{i}}\) in the \(z\) polarized phase close to the critical point, \((h/J)_{\mathrm{i}} = 0.4\);
  (c)~\(\hat{H}_{\mathrm{i}}\) in the \(x\) N\'eel phase, \((h/J)_{\mathrm{i}} = 0.25\);
  (d)~\(\hat{H}_{\mathrm{i}}\) deep in the \(x\) N\'eel phase, \((h/J)_{\mathrm{i}} = 0.1\).
  The dashed blue and solid green lines  are algebraic fits to propagating local extrema and to the CE respectively.
}
\label{local_quench_quasi_local_regime}
\end{figure}

Figure~\ref{local_quench_quasi_local_regime} shows results for both the \(G_x(R,t)\) (upper row) and the \(G_z(R,t)\) (lower panel) correlation functions in the quasi-local regime, \(\alpha=1.7\). Form left to right, we quench from decreasing values of the initial parameter \((h/J)_\textrm{i}\), ranging from the \(z\)-polarized phase to the \(x\)-N\'eel phase. The post-quench parameter \((h/J)_\textrm{f}\) is fixed to a value corresponding to the \(z\)-polarized phase.
When changing the initial value \((h/J)_\textrm{i}\) towards the N\'eel phase, we find that an additional staggered structure, reminiscent of the N\'eel ordering, appears. For the \(G_x(R,t)\) correlation function, it rapidly dominates the signal and it becomes difficult to identify propagating local maxima and a correlation edge. In contrast, the signal is clearer for the \(G_z(R,t)\) correlation function. While a checkerboard-like structure appears, it is still possible to identify the local maxima and the correlation edge. We find that both are algebraic with dynamical exponents \(\betam\) and \(\betaCE\) that do not significantly change with the strength of the quench.

\begin{figure}[!ht]
\centering
\begin{tabular}{c}
\includegraphics[width = \columnwidth]{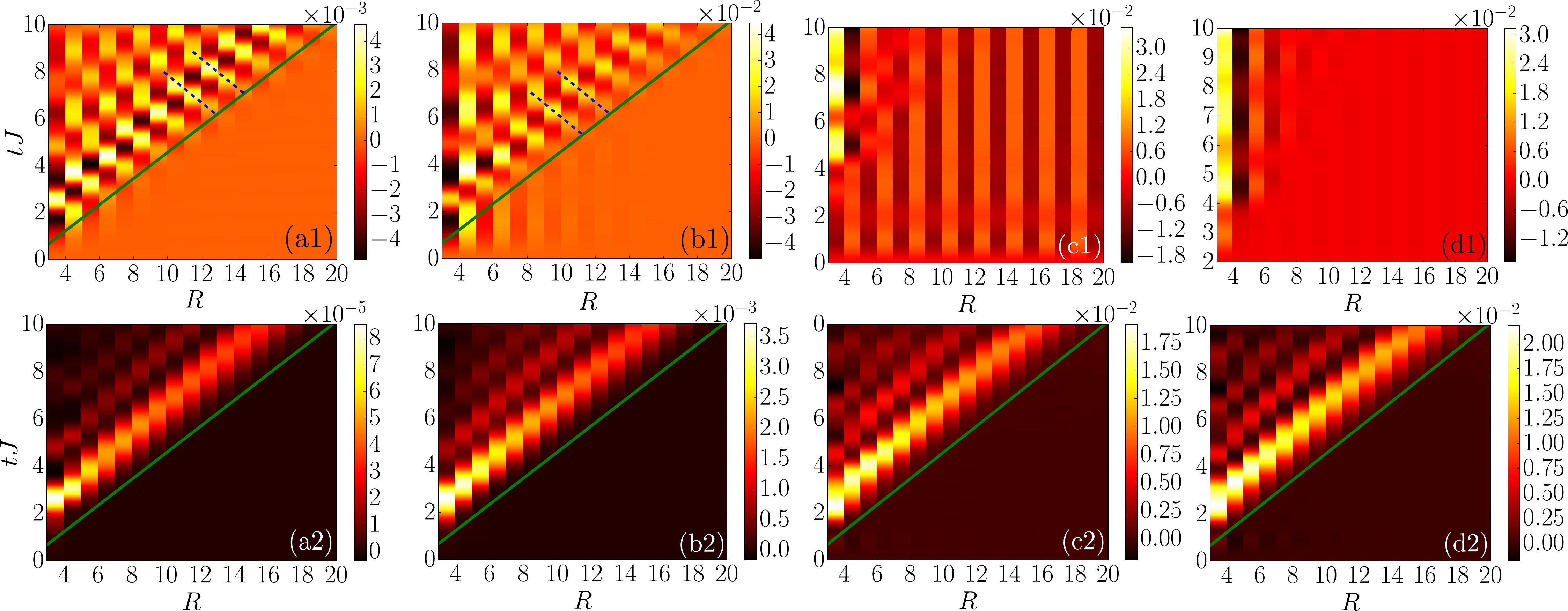}
\end{tabular}
\caption{Spreading of the \(G_x\) (upper row) and \(G_z\) (lower row) spin correlations in the local regime, \(\alpha = 3\) (linear scale).
The quench is performed from an increasing value of the initial parameter \((h/J)_{\mathrm{i}}\), ranging from the \(z\) polarized phase to the \(x\)-N\'eel phase,
to the fixed value \((h/J)_{\mathrm{f}} = 1\), corresponding to the \(z\)-polarized phase.
(a)~\(\hat{H}_{\mathrm{i}}\) deep in the \(z\)-polarized phase, \((h/J)_{\mathrm{i}} = 0.9\);
(b)~\(\hat{H}_{\mathrm{i}}\) in the \(z\) polarized phase close to the critical point, \((h/J)_{\mathrm{i}} = 0.59\);
(c)~\(\hat{H}_{\mathrm{i}}\) in the \(x\) N\'eel phase, \((h/J)_{\mathrm{i}} = 0.25\);
(d)~\(\hat{H}_{\mathrm{i}}\) deep in the \(x\) N\'eel phase, \((h/J)_{\mathrm{i}} = 0.1\).
The dashed blue and solid green lines  are linear fits to propagating local extrema and to the CE respectively.}
\label{local_quench_local_regime}
\end{figure}

Figure~\ref{local_quench_local_regime} shows the same calculations, now performed in the local regime, \(\alpha=3\). Similar conclusions as for the quasi-local regime hold. The \(G_x(R,t)\) correlation function is rapidly blurred. In contrast, the \(G_z(R,t)\) is more robust and allows us to fit the dynamical exponents \(\betam\) and \(\betaCE\), yielding values nearly independent of \((h/J)_\textrm{i}\). In fact, for the \(G_z(R,t)\) correlation function, we find no significant impact of increasing the strength of the quench.

\section{Scaling laws for \(\mathbf{\langle S^z_R(t)\rangle}\) after a local quench} \label{local_mag}
The local magnetization \(\langle \hat{S}^z_R(t) \rangle\) can be computed analytically for a local quench in the \(z\)-polarized phase where \(h \gg J\), using an approach similar to that presented in Sec.~\ref{LSWT}.
Starting from the initial  state
\(\ket{\Psi_0} \propto \hat{S}^-_{0} \ket{\Psi_\textrm{\tiny GS}}\) obtained by lowering the central (\(R=0\)) spin of the ground state (which is close to the product state \(\ket{\uparrow ... \uparrow \downarrow \uparrow ... \uparrow}\)),
and applying both the Holstein-Primakoff transformation and the bosonic Bogoliubov approach,
we find
\begin{align}
 & 1/2 - \langle \hat{S}^z_{R}(t) \rangle \simeq \left| \int_{-\pi}^{\pi} \frac{\mathrm{d}k}{2\pi} \mathcal{F}_1(k) \frac{e^{i(kR + E_kt)}+ e^{-i(kR - E_kt)}}{2}
 \right|^2 + \left| \int_{-\pi}^{\pi} \frac{\mathrm{d}k}{2\pi} \mathcal{F}_2(k) \frac{e^{i(kR + E_kt)}+ e^{-i(kR - E_kt)}}{2}
 \right|^2.
 \label{eq_expec_val}
\end{align}
The quantities \(\mathcal{F}_1(k)\) and \(\mathcal{F}_2(k)\) are two different quasi-momentum-dependent amplitude functions, which read as
\begin{equation}
\mathcal{F}_1(k) = \frac{1}{2}\left( \frac{\mathcal{A}_k}{E_k} +1 \right) = \frac{1}{2} \left( \frac{2h + JP_{\alpha}(k)}{2\sqrt{h[h+JP_{\alpha}(k)]}} + 1\right)
\qquad \textrm{and} \qquad
\mathcal{F}_2(k) = -\frac{\mathcal{B}_k}{2E_k} = -\frac{J P_{\alpha}(k)}{4\sqrt{h[h+JP_{\alpha}(k)]}}.
\label{amp_fct_local_spin}
\end{equation}
For large transverse fields, \(h \gg J\), we find $\mathcal{F}_1(k) \simeq 1 \gg \mathcal{F}_2(k)
\sim J/h$ and the second term in Eq.~(\ref{eq_expec_val}) may be neglected.

\bigskip
\paragraph{Quasi-local regime~--~}
In the quasi-local regime, owing to the infrared divergence of the quasi-particle velocities, the spreading of the local magnetization (as well as that of correlations) is dominated by the Fourier components with \(k \rightarrow 0\), see Eq.~(\ref{eq_expec_val}). There, we may write \(E_k \simeq \Delta - c \vert k\vert^z\), where \(\Delta = 2\sqrt{h[h+JP_\alpha(0)]}\) is the gap and \(z=\alpha-1\) is the dynamical exponent, see also Sec.~\ref{SM:QLregAnalytics}. Inserting this expression into Eq.~(\ref{eq_expec_val}), we find that the gap \(\Delta\) may be factorized, taken out of the integrals, and then vanishes owing to the square moduli. It explains the irrelevance of the gap for the spreading of the local magnetization in the quasi-local regime. Applying the stationary phase approximation to evaluate the integrals along the lines of Ref.~\cite{cevolani2018}, we then find that the spreading of \(\langle \hat{S}^z_{R}(t) \rangle\) is super-ballistic, with \(\betam = 1 -z\).
This is consistent with the TDVP results shown in the main paper, see Fig.~\ref{fig:numerics2}(b) of the main paper.

\bigskip
\paragraph{Local regime~--~}
In the local regime, both the quasi-particle energy \(E_k\) and group velocity \(\Vg(k)\) are finite over the whole Brillouin zone. 
The spin edge is thus determined by the maximum group velocity, \(Vg(k^*)\).
The calculation is similar to that outlined for the spreading of the spin correlation in Sec.~\ref{SM:LocRegAnalytics}.
The only relevant difference is that the phase factor for the spreading of the spin correlations has an extra factor of \(2\), which is absent for the local magnetization, see Eq.~(\ref{gx_analy}) versus Eq.~(\ref{eq_expec_val}). As a result, the spin edge propagates at the velocity \(\VSE=\Vg(k^*)\), while the correlation edge propagates at the velocity \(\VCE=2\Vg(k^*)\).
This is consistent with the TDVP results, see Fig.~\ref{fig:numerics2}(c) of the main paper~\footnote{Note that a different value of \(h/J\) is used in the two figures and the values of \(\VCE\) in Fig.~\ref{fig:numerics2}(c) and \(\VSE\) in Fig.~\ref{fig:numerics3}(c) in the main paper are not related.}.

More precisely, applying the stationary phase approximation, we find
\begin{align}
 & 1/2 - \langle \hat{S}^z_{R}(t) \rangle \sim \left|\frac{\mathcal{F}_1(k^*)}{(|\partial^2_k E_{k^*}|t)^{1/2}} [\cos(k^* R
 -E_{k^*} t + \phi) - i \sin(k^* R - E_{k^*} t + \phi) ]\right|^2 
 \label{expec_spa}
\end{align}
where \(\phi = - (\pi/4)\mathrm{sgn}(\partial_k^2 E_{k_\mathrm{sp}}t)\) is a constant phase term.
The local magnetization is thus the sum of two terms, associated to the \(\cos\) and \(\sin\) terms.
Each has local maxima that propagate at the velocity \(\Vphi(k^*)\). The two terms are, however, in phase opposition, which mutually cancel the local maxima. We hence expect a single structure in the vicinity of the spin edge, consistently with the TDVP results. 
A similar effect has been found for density correlations deep in the Mott insulator phase of the Bose-Hubbard model~\cite{despres2019}.
This effect is also expected in the quasi-local regime but the extinction of the local maxima happens on very large distances and in the TDVP we still observe them up to the largest accessible times and distances, see Fig.~\ref{fig:SpinCorr} of the main paper.

\section{Density matrix and entanglement entropies after a local quench}
\label{app:density_matrix_local}

In this section we discuss analytic and numerical results for the entanglement entropy following a local quench, using linear spin wave theory (LSWT).

\subsection{Spin wave analysis}
The initial state prepared by the local quench is obtained from the ground state (GS) by an up-down spin flip at the central lattice site. The time dependence of the system's state is given by
\begin{equation}
  \ket{\Psi_{0}(t)} = \tilde{\mathcal{N}} \hat{S}^-_0 (t) \ket{\Psi_{\mathrm{GS}}} = \tilde{\mathcal{N}} \hat{a}^\dagger_0(t) \ket{0_b} = \tilde{\mathcal{N}} \sum_{k=-\pi}^{\pi} u_k \mathrm{e}^{\mathrm{i} E_k t}\, \hat{b}_k^\dagger(t=0) \ket{0_b} \,,
\label{eq.initial_state}
\end{equation}
where \(\hat{S}^-_0 = S^x_0 - i S^y_0\) is the spin lowering operator at the central site (labelled by \(j=0\)), \( \hat{a}_0^{\dagger}\) is the bosonic Holstein-Primakoff creation operator [see Eq.~(\ref{HP})], \(\hat{b}_k(0)\) is the Bogoliubov quasi-particle annihilation operator~\cite{bogoliubov1947}, \(u_k = \mathrm{sign}(\mathcal{A}_k) \sqrt{\frac{1}{2} \left(\frac{\abs{\mathcal{A}_k}}{E_k} + 1 \right)}\),
\(\ket{0_b}\) is the quasi-particle vacuum, and \(\tilde{\mathcal{N}}=\left(\sum_k u_k^2\right)^{-1/2}\) stands for the normalization of the state.
Crucially, for \(h/J \gg 1\), the initial state is a superposition of single-particle excitations,
which allows us to proceed analytically.
Defining \(S\) as the entire system and
 \(A\) and \(B\) as complementary subsystems, we may decompose the initial state into the cases where the excitation is in either subsystem \(A\) or \(B\) as \( \sum_{m \in {S}} \ket{1_m} = \sum_{m \in A} \ket{1_m} \otimes \ket{0_{B}} + \sum_{m \in B}\ket{0_{A}} \otimes \ket{1_m}\), where \(\ket{0_{{A},{B}}}\) is the unique vacuum of the subsystems \(A\) and \(B\), respectively. The state of the entire system thus reads as
\begin{align}
  \ket{\Psi_0(t)} &= \tilde{\mathcal{N}} \sum_{m \in B} \sum_{k=-\pi}^{\pi} u_k \mathrm{e}^{\mathrm{i} \left( E_k t + m k\right) } \ket{0_A} \otimes \ket{1_m} + \tilde{\mathcal{N}}\sum_{m \in A} \sum_{k=-\pi}^{\pi} u_k \mathrm{e}^{\mathrm{i} \left( E_k t + m k\right) } \ket{1_m} \otimes \ket{0_B} \\
  &\coloneqq \sqrt{\lambda_1} \mathrm{e}^{\mathrm{i} \theta_1} \ket{0_A} \otimes \ket{\chi_B} + \sqrt{\lambda_2} \mathrm{e}^{\mathrm{i} \theta_2} \ket{\chi_A} \otimes \ket{0_B} \label{eq:lambda2_def} \,,
\end{align}
where \(\lambda_1\) and \(\lambda_2\) are real-valued numbers such that \(\lambda_1 + \lambda_2 = 1\), and \(\theta_1\) and \(\theta_2\) are some phases.
The density matrix of the entire system thus contains four terms,
\begin{align}
  \rho(t) &= \lambda_1 \ketbra{0_A} \otimes \ketbra{\chi_B}{\chi_B} + \lambda_2 \ketbra{\chi_A}\otimes \ketbra{0_B} \nonumber \\ 
  &\quad + \sqrt{\lambda_1 \lambda_2}\mathrm{e}^{i(\theta_1-\theta_2)} \ketbra{0_A}{\chi_A} \otimes \ketbra{\chi_B}{0_B} + \sqrt{\lambda_1 \lambda_2}\mathrm{e}^{i(\theta_2-\theta_1)} \ketbra{\chi_A}{0_A} \otimes \ketbra{0_B}{\chi_B} \,,
\end{align}
while the reduced density matrix of subsystem \(A\) contains two terms,
\begin{align}\label{eq:reduced_density_matrix}
  \rho_A(t) &= \tr_B(\rho(t)) = \lambda_1(t) \ketbra{0_A} + \lambda_2(t) \ketbra{\chi_A} \,.
\end{align}

In the limit of \(h/J\gg1\) the Bogoliubov coefficient approaches unity, \(u_k \simeq 1 \). 
Since the two eigenvalues are constrained to add to unity and have the same functional form, we may without loss of generality restrict our analysis to one of them.
The \(t\rightarrow \infty\,, R\rightarrow \infty\) behavior of the second eigenvalue of the reduced density matrix in the infinite volume limit using the stationary phase approximation is given by
\begin{align}
  \lambda_2 &\simeq {\tilde{\mathcal{N}}^2} \sum_{m\in A} \abs{\sum_{k=-\pi}^{\pi}\mathrm{e}^{\mathrm{i}\left(E_{k}t + mk \right)}}^2 \label{eq:lambda_2} \overset{N\rightarrow \infty}{\propto} \tilde{\mathcal{N}}^2  \,
  \sum_{m \in A} \frac{1}{t \abs{ \partial^2_{k} E_{k_{\mathrm{sp}} } }} \cos^2\!\big[ E(k_{\mathrm{sp}})t - k_{\mathrm{sp}}m + \phi \big]
\end{align}
where \(k_{sp}\) is defined by the stationary phase condition \(\Vg(k_{\mathrm{sp}}) = m/t \) with \(\Vg(k) = \partial E_k/\partial k\).

\subsection{Entanglement Entropies}

In order to compute the entanglement entropies from the reduced density matrix,
we now consider the partitions \(A = \left\{ R,R+1,\ldots, N/2 \right\}\) and \(B = \left\{ -N/2,-N/2+1,\ldots, R -1\right\}\) and look for an analytic expression for its eigenvalues.

\bigskip
\paragraph{Quasi-local regime~--~}
The spectrum in the quasi-local regime is bounded but its first derivative \(V_g(k)\) has an infrared divergence (\(k_{sp} \to 0\)), as \(V_g(k) \sim 1/k^{2-\alpha}\), which dominate the eigenvalues of the reduced density matrix.
Inserting this into \cref{eq:lambda_2}, we find that the second eigenvalue can be approximated in the infinite volume limit by,
\begin{align}
  \lambda_2 &\propto \tilde{\mathcal{N}}^2  \,
   t^{\frac{1}{1-z}} \sum_{m=0}^\infty \frac{1}{(m+R)^{\frac{z-2}{z-1}}} \cos^2\left[\xi_z \left( \frac{t }{(m+R)^z} \right)^{\frac{1}{1-z}} + \phi \right]
   \label{eq.lambda2approx}
\end{align}
with \(\xi_z = (cz)^{\frac{1}{1-z}} \left( 1 + c^{z+1}z^z \right)\).
In contrast with the earlier study of spin wave excitations for magnetic properties (see for instance Secs.~\ref{LSWT} and \ref{local_mag}), here the site \(R\) is contained in every term of the sum over subsystem \(A\), and we cannot extract the entanglement edge without considering every term in the sum. Moreover, this means that we cannot \emph{a priori} simply look at the scaling of the coefficient without taking into account the oscillatory cosine term. However, in both our TDVP and LSWT numerical results, we find no significant internal oscillations of the R\'enyi entanglement entropies, implying that the oscillatory term does not contribute significantly to the observed behavior of the entanglement edge. Additionally, we find the frequency \(\xi_z\) to be very large such that the cosine term varies rapidly compared to the algebraically decaying amplitude.
We can then safely substitute the square cosine function by its average value, \(1/2\),
allowing us to analytically perform the sum in Eq.~(\ref{eq.lambda2approx}).
We find
\begin{align}
  \lambda_2 & \propto \tilde{\mathcal{N}}^2 \frac{2\pi(cz)^{\kappa}}{\abs{cz(z-1)}} t^{\frac{\kappa}{2-z}} \zeta\left(\kappa, R\right) \,,
\end{align}
where \(\zeta(s,q) = \sum_{m=0}^\infty {(m+q)^{-s}}\) is the Hurwitz zeta function~\cite{Hurwitz1932, Magnus1966} and \(\kappa = \frac{z-2}{z-1} = \frac{\alpha-3}{\alpha-2}\). Using the series representation of the Hurwitz zeta function in the limit of large \(R\) and positive \(\kappa\), this may be written as
\begin{align}
  \lambda_2 &= \tilde{\mathcal{N}}^2 \frac{2\pi(cz)^{\kappa}}{\abs{cz(z-1)}} t^{\frac{\kappa}{2-z}} \left[ \frac{R^{1-\kappa}}{\kappa-1} + \frac{R^{-\kappa}}{2} + \sum_{j=1}^\infty \frac{B_{2j}}{(2j)!} \frac{\Gamma(\kappa + 2j -1)}{\Gamma(\kappa)} R^{1-\kappa-2j} \right] \,,
\label{eq.lambda2-series}
\end{align}
where \(B_n\) are the Bernoulli numbers, and \(\Gamma(s) = \int_0^\infty \mathrm{e}^{-u} u^{s-1} \dd{u}\) is the gamma function~\cite{Magnus1966}. In the quasi-local regime, \(\kappa > 2\) and so in the limit of large \(R\) we may neglect the second and third term which contain contributions of the order equal to \(R^{-\kappa}\) and higher, and only consider the leading term of order \(R^{1-\kappa}\).
It yields
\begin{align}
  \lambda_2 &\approx \tilde{\mathcal{N}}^2 \frac{2\pi(cz)^{\kappa}}{\abs{cz(z-1)}} \frac{1}{\kappa-1} \left(\frac{t}{R}\right)^{\frac{1}{1-z}} \,
  \label{eq.approx-lambda2}
\end{align}
which scales in \(R\) and \(t\) with the same power. Note that the stationary phase approximation is not valid when taking the limit \(t\to\infty\) alone, and the above equation does not hold in this limit.
Consequently, any analytic function of \(\lambda_2\), particularly R\'enyi entropies, will scale in \(R\) and \(t\) with the same power, so that
\begin{align}
  \mathcal{S}_{n}(R,t) \simeq \frac{1}{1-n} \log\!\big\{\lambda_2\!\left({R}/{t}\right)^n + \big[1-\lambda_2\!\left({R}/{t}\right)\big]^n\big\} \, .
\end{align}
This yields a dynamical exponent of the EE of unity, \(\beta^{n}_{EE} = 1 \ \forall \, n\), in close agreement with our numerical results using both TDVP and LSWT.
The additional finite-size corrections neglected in moving from Eq.~(\ref{eq.lambda2-series}) to Eq.~(\ref{eq.approx-lambda2}) are likely responsible for the deviation from ballistic behavior seen in our simulations on small system sizes.

\bigskip
\paragraph{Local regime~--~}
The calculation in the local regime \(\alpha > 2\) proceeds similarly. However, both the quasi-particle energy \(E_k\) and group velocity \(\Vg(k)\) are finite over the whole Brillouin zone such that there exists a quasi-momentum \(k^\star\) where the group velocity is maximum, \(\max_k V_g(k) = V_g(k^\star)\). The stationary phase condition in the local regime thus has only a solution for \(m/t \leq V_g(k^\star)\). The eigenvalues of the reduced density matrix are then dominated by the momenta \(k \sim k_{\mathrm{sp}} = k^\star\), which is independent of \(R\) and \(t\), and thus scale ballistically. Finally, any analytic function of the eigenvalues will also scale ballistically.

\section{R\'enyi entropies after a local quench}\label{sec:ent_entropies}

In this section we present complementary numerical results for different R\'enyi entropies in the local and quasi-local regimes obtained via TDVP simulations and LSWT calculations. 

\begin{figure}
\centering
\begin{minipage}{0.3\textwidth}
  \includegraphics[width=\textwidth]{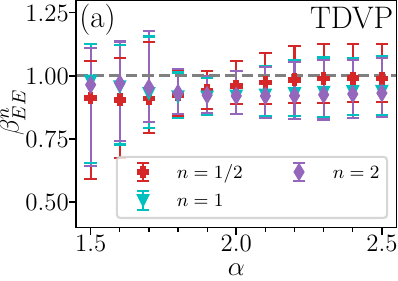} 
  \includegraphics[width=\textwidth]{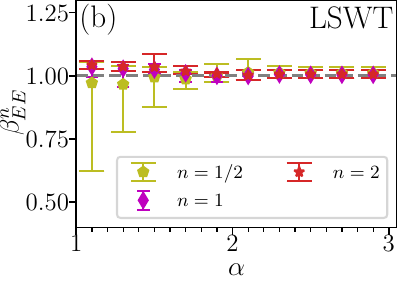}
\end{minipage}
\begin{minipage}{0.69\textwidth}
  \includegraphics[width=0.32\textwidth]{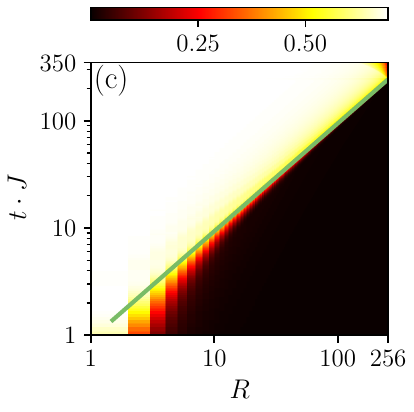}
 \includegraphics[width=0.32\textwidth]{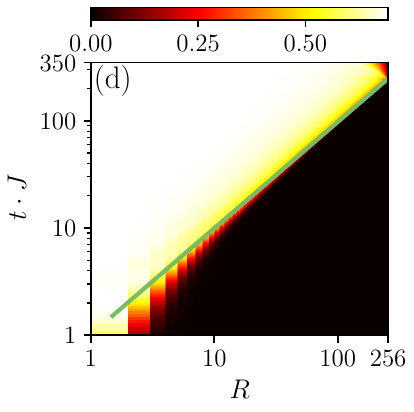}
\includegraphics[width=0.32\textwidth]{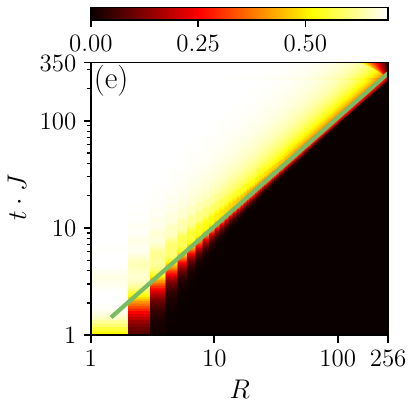}
\includegraphics[width=0.32\textwidth]{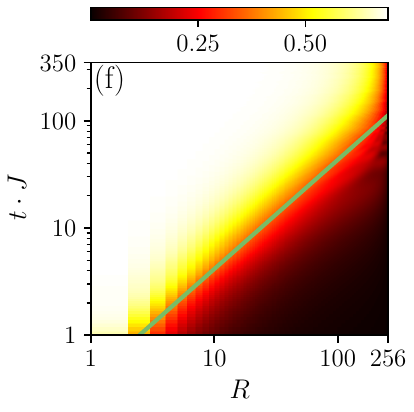}
\includegraphics[width=0.32\textwidth]{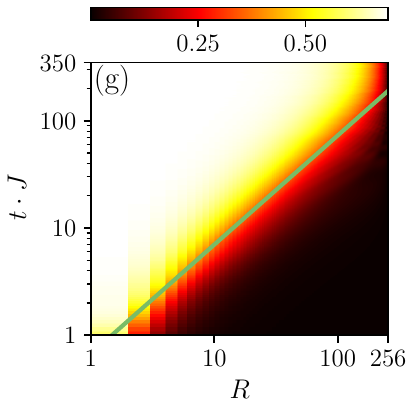}
\includegraphics[width=0.32\textwidth]{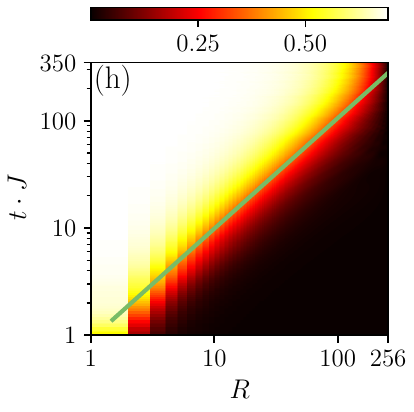}
\end{minipage}
\caption{
  EE after a local quench in the \(z\)-polarized phase (\(h/J = 50\)).
  (a) TDVP (\(N=96\)): dynamical exponent \(\beta_{EE}^n\) of the three R\'enyi entropies over different values of \(\alpha\).
  (b) LSWT (\(N=512\)): dynamical exponent \(\beta_{EE}^n\) of the three R\'enyi entropies over different values of \(\alpha\).
(c)--(h) LSWT (\(N=512\)): Space-time behavior of the three R\'enyi entropies \(\mathcal{S}_n(R,t)\) considered in the local regime (\(\alpha=2.5\)) with \(n=1/2\) (c), \(n=1\) (d), and \(n=2\) (e) as well as in the quasi-local regime (\(\alpha=1.5\)) with \(n=1/2\) (f), \(n=1\) (g), and \(n=2\) (h). The solid green lines represent power law fits to the entropy edges with dynamical exponents, (c) \( \beta_{EE}^{n=1/2} = 1.006 \pm 0.001\), (d) \(\beta_{EE}^{n=1} = 0.9907 \pm 0.0002\), (e) \(\beta_{EE}^{n=2} = 1.014 \pm 0.002\), and (f) \(\beta_{EE}^{n=1/2} = 1.018 \pm 0.003 \), (g) \(\beta_{EE}^{n=1} = 1.018 \pm 0.001 \), (g) \(\beta_{EE}^{n=2} = 1.031 \pm 0.002 \).
}
\label{fig:local_quench_entropy-beta}
\end{figure}

In \cref{fig:local_quench_entropy-beta}(a) and (b), 
the counterparts of Fig.~\ref{fig:numerics3}(b) and (c) in the main text, dynamical exponents \(\beta_{EE}^n\) are shown for the different values of the R\'enyi parameter considered (\(n=1/2\), \(n=1\), \(n=2\)) obtained via TDVP and LSWT with system size \(N=96\) and \(N=512\), respectively.
The dynamical exponent of the EE remains close to unity almost independently of the order of the R\'enyi entropy, both in TDVP and LSWT.
Each of the \(\mathcal{S}_n(R,t)\) considered displays a linear causal structure fully characterized by an entanglement edge (EE) and its behavior is very similar to \cref{fig:numerics3}(a).

We have also computed the entanglement entropy on much larger system sizes using LSWT. This allows us to obtain a better estimate of the spreading exponent \(\betaEE{n}\) in the thermodynamic limit. The results are shown in Fig.~\ref{fig:local_quench_entropy-beta}(c)--(h) for three values of the R\'enyi order \(n=1/2\), \(1\), and \(2\) and two values of \(\alpha\) for a chain of length \(N=512\). The results of the fits to the EE are shown in \cref{fig:local_quench_entropy-beta}(b), where we see that the error bars of the von Neumann entanglement entropy are much smaller and give a more precise indication of ballistic spreading of the EE.

The determination of the EE emerges as a challenge, due to the absence of a sharply defined edge in the quasi-local regime.
In practice we chose a large range and average over it to find a value for \(\beta_{EE,n}\) in Fig.~\ref{fig:numerics3}(b) and (c). The threshold ranges for all R\'enyi entropies \(\mathcal{S}_{n}\) obtained from TDVP and LSWT in all system sizes is \(\epsilon \in [20\%,80\%]\). The associated error is taken to be the difference between the maximal and minimal value for the dynamical exponent within this range of thresholds.

\end{document}